# Demonstrating the Suitability of Neuromorphic, Event-Based, Dynamic Vision Sensors for In Process Monitoring of Metallic Additive Manufacturing and Welding

**David Mascareñas[1]\*, Andre Green[2], Ashlee Liao[3], Michael Torrez[4], Alessandro Cattaneo[4], Amber Black[5], John Bernardin[6], Garrett Kenyon[7]**

[1]  Los Alamos National Laboratory; dmascarenas@lanl.gov
[2]  Los Alamos National Laboratory; andre_green@lanl.gov
[3]  John Hopkins Applied Physics Laboratory (Current- however during execution of work Ashlee Liao was affiliated with Los Alamos National Laboratory); liaoashlee@gmail.com
[4]  Los Alamos National Laboratory; torrez@lanl.gov
[5]  Los Alamos National Laboratory; cattaneo@lanl.gov
[6]  Los Alamos National Laboratory; anblack@lanl.gov
[7]  Los Alamos National Laboratory; bernardin@lanl.gov
[8]  Los Alamos National Laboratory; gkenyon@lanl.gov

\*  Correspondence: dmascarenas@lanl.gov ;

**Abstract:** We demonstrate the suitability of high dynamic range, high-speed, neuromorphic event-based, dynamic vision sensors for metallic additive manufacturing and welding for in-process monitoring applications. In-process monitoring to enable quality control of mission critical components produced using metallic additive manufacturing is of high interest. However, the extreme light environment and high speed dynamics of metallic melt pools have made this a difficult environment in which to make measurements. Event-based sensing is an alternative measurement paradigm where data is only transmitted/recorded when a measured quantity exceeds a threshold resolution. The result is that event-based sensors consume less power and less memory/bandwidth, and they operate across a wide range of timescales and dynamic ranges. Event-driven driven imagers stand out from conventional imager technology in that they have a very high dynamic range of approximately 120 dB. Conventional 8 bit imagers only have a dynamic range of about 48 dB. This high dynamic range makes them a good candidate for monitoring manufacturing processes that feature high intensity light sources/generation such as metallic additive manufacturing and welding. In addition event based imagers are able to capture data at timescales on the order of 100 μs, which makes them attractive to capturing fast dynamics in a metallic melt pool. In this work we demonstrate that event-driven imagers have been shown to be able to observe tungsten inert gas (TIG) and laser welding melt pools. The results of this effort suggest that with additional engineering effort, neuromorphic event imagers should be capable of 3D geometry measurements of the melt pool, and anomaly detection/classification/prediction.

LA-UR-24-25981

## 1. Introduction

In process monitoring is an important problem that, if solved, would help allow the use of additive manufacturing (AM) for mission-critical metallic components. In-process monitoring has been proposed as a solution for reducing the time needed for non-destructive evaluation (NDE) by identifying potentially problematic areas of interest for further inspection in the completed part. Unfortunately, there are a number of challenges that make in-process monitoring difficult for the high-temperature, high light environment presented by metallic additive manufacturing. The first problem is that the pixels of conventional imagers become saturated and typically do not have the dynamic range needed to observe melt pool processes (Figure 1). Pixel saturation occurs when the total amount of charge at a pixel exceeds the full well capacity of that pixel [1], [2]. High dynamic range imagers exist, but they typically only have a frame rate on the order of tens of Hertz which is not suitable for the fast dynamics occurring in a melt pool on the order of hundreds of Hertz [3]. High speed laser illumination can be used in combination with high-speed cameras [4], but these approaches are expensive and result in extremely large amounts of data not suitable for in-process monitoring or forming digital twins of additively manufactured components. To solve these issues, we propose the use of event-based imagers. Event-based sensing is an alternative measurement paradigm where data is only transmitted/recorded when the measured quantity exceeds a threshold resolution [5]. Traditionally, event imagers have been designed to detect changes in log light intensity. The pixels traditionally detect changes in log light intensity independently of one another. The threshold is set according to the parameters of a photoreceptor circuit [5]. Traditional event imagers have a nominal, specified log change detection threshold that is the same across all the pixels of the event imager at any given time. In reality, there is some variation present in event-imager pixel electronics [5]. Traditional event imagers also feature programmable bias current generators to perform tasks such as reduce power consumption and set bandwidth [5]. The result is that event-based sensors consume less power and less memory/bandwidth, and they operate across a wide range of timescales and dynamic ranges [5]. Furthermore, these imagers require approximately 35 times less memory to store data comparable to that of a conventional imager. In future work, we envision that neuromorphic sensors can play a role in the next generation of On Machine Inspection (OMI) [6]. Currently OMI is lacking in the AM world and if we are to qualify a machine and process, we need efficient, reliable and repeatable OMI methods and some kind of Figure of Merit (FOM) on the build process to establish a Go/NoGo for a built part. Some potential OMI applications include:

1. Linking the transient melt pool geometry (shape and fluctuations) and temperature to a TRUCHAS [7], [8], [9] code simulation to better understand the cooling, gain and phase development of the metal during the printing process.
2. Defect detection in the build process (keyholing, blowout, under heating, etc.)
3. Check for consistency in repeated build layer height or from repeated builds of identical parts. The data collected by a neuromorphic OMI system would capture the build histories of parts. This data could be analyzed to determine if repeated builds are similar or different and by how much, and if the differences are significant or not (part of a machine/process qualification). The OMI build history data could also be analyzed to find relationships with post build/CT scanning analysis techniques as part of a general qualification method for metal AM.

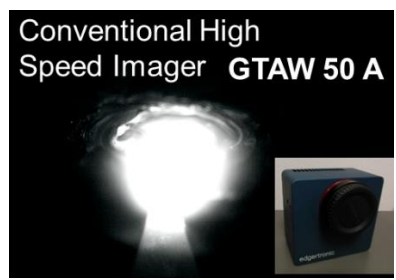

Figure 1 A conventional high-speed imager suffers from saturation during arc welding processes.

### 1.1. Overview of Neuromorphic Sensor Operation

Event-based sensing is an alternative measurement paradigm where data is only transmitted/recorded when a threshold resolution is exceeded ( *[5]*). Event-based sensors perform the detection of light intensity changes at the individual pixel level in hardware *[5]*. Data is only reported from the imager if the light intensity level at a given pixel changes beyond the threshold. Furthermore, only the pixels whose light intensity value has changed report data. The data for each event consists of an event timestamp that typically has tens of microseconds temporal resolution *[10]* as well as the x-y position of the pixel that detected the event and a polarity value which indicates if the pixel detected an increase or decrease in light intensity. If the pixel event data is arranged into the form of a photograph, the resulting data looks like a conventional image that has been processed using an edge detection algorithm.

This strategy used by neuromorphic technologies is inspired by the efficient sensing and data processing done by biological nervous systems *[11]*, *[12]*. The result of this detection-at-the-pixel approach, is that event-based sensors consume less power and less memory/bandwidth, and they operate across a wide range of timescales and dynamic ranges *[5]*.

At this time a number of commercially available neuromorphic imagers for the visible light range are available including those from iniVation *[10]*, Synsense *[13]*, and Prophesee *[14]*. In 2023 Quantum Imaging has announced that they will offer a synchronous event-based imaging sensor that will operate in the short wave infrared (SWIR) wavelength range *[15]*.

## 2. Materials and Methods

### 2.1. Gas Tungsten Arc Welding Experiment

To evaluate the suitability of event-based imagers for monitoring melt pools associated with additive manufacturing and welding, a DVS 240 C event based imager was used to observe a welding process (*Figure 2*). A weld shade was placed in the optical axis of the imager optics in order to reduce the total amount of light observed by the imager. A gas tungsten arc welding (GTAW) torch was used to generate a melt pool on a pipe structure. An example of the events captured in the process of this experiment can be seen in *Figure 3* and a frame generated from the events can be seen in *Figure 4*. From this data the outline of the melt pool can be made out clearly. It is also possible to observe the outline of the tungsten electrode. In the course of these experiments we could also observe small anomalies that generated large numbers of events relative to the rest of the melt pool that would move in coherent ways through out the pool in time. An example of these anomalies can be seen in *Figure 4*. At this time the team hypothesizes that these anomalies are associated with contaminants in the melt pool that possess higher emissivity than the rest of the melt pool. The current hypothesis is that some oil entered the melt pool, however, this hypothesis will need to be explored further in future work. These results demonstrate that event-based imagers have high potential for observing melt pools generated by electric arcs.

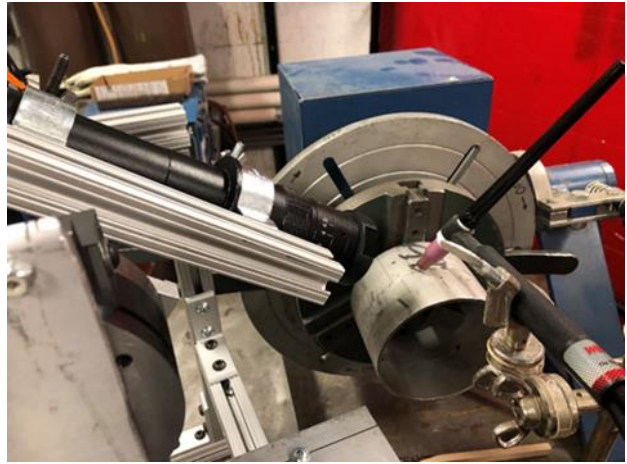

*Figure 2    Test setup associated with a DVS 240 C event-based imager observing a gas tungsten arc welding process.    In this photo the welding torch is held by a clamp.    In the course of this effort, data was collected both for the torch held by a clamp as well as for the case of the torch being held manually in a welder's hand.*

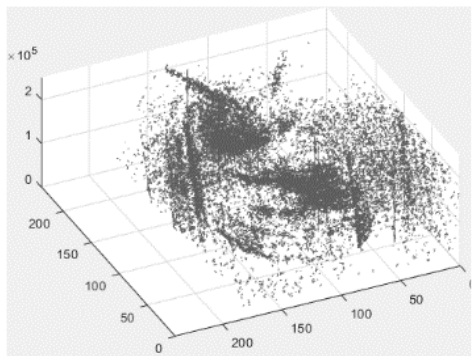

*Figure 3 Event data captured from the GTAW melt pool.    The vertical axis is associated with time in microseconds.    The two axis at the bottom are the x and y pixel indices.*

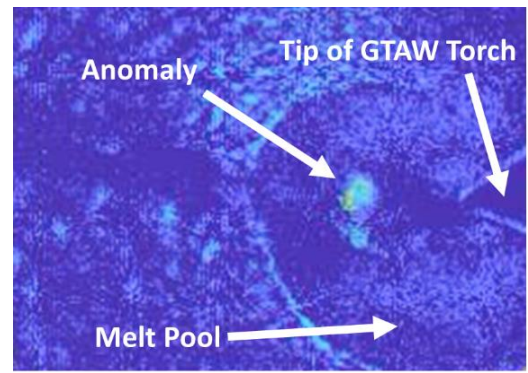

*Figure 4 Event-based data formed into a frame representation for the event imager observing the GTAW/TIG process.    In this case an anomaly can be observed in the melt pool.*

### 2.2 Laser Welding Experiment

Next we consider the case of using event-based imagers to observe melt pools generated by lasers.    The laser welding setup is shown in *Figure 5*.    A Davis 346 event based imager was used to observe the melt pool.

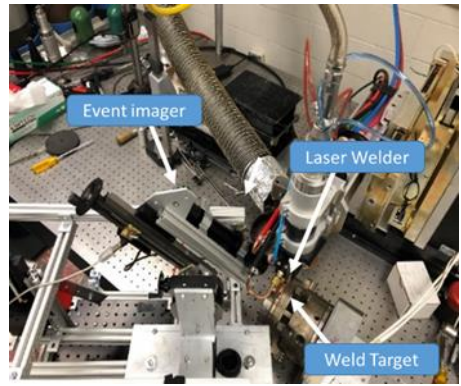

*Figure 5 Laser welder being observed with an event-based imager.*

Figure 6 below shows some examples of event-based imager data of the laser welding process the team captured at different points in time. From these images the change in aspect ratio of the keyhole can clearly be observed at different points in time. Knowing information about the geometry of melt pools can be important for in-process monitoring applications [16]. We anticipate that the predictive models we create may help aid in the discovery of new melt pool phenomena that predict changes in geometry that may be indicative of the onset of defects.

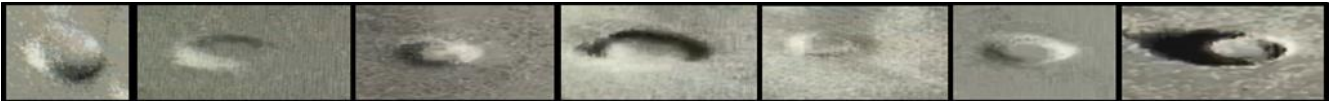

*Figure 6 Examples of event-based imagery captured from a laser welding keyhole. Note how the aspect ratio of the keyhole changes for different points in time.*

## 3. Results

### 3.1. Adaptive sampling and memory saving properties

We now demonstrate the ability of event-based imagers to perform adaptive sampling. In the context of event imagers, adaptive sampling refers to the ability of event imagers to only detect/generate events at the timescales relevant to the phenomena being observed at each pixel independent of one another. To demonstrate this capability we consider the case of an event-based imager being used to observe a popping balloon. In this work a DVXplorer event-based imager was used to observe a biaxially oriented polyethylene terephthalate (BoPET) (trade name Mylar) balloon with a checkerboard pattern print. The event-based imager with high speed camera can be seen in Figure 7. Figure 8 shows frames generated from the event-based imager data along with a plot that indicates with a vertical dashed line what the cumulative ratio of the total number of events associated with the popping event is at a given point in time. The balloon was observed for a total of about 10 seconds and the popping event occurred at about 3 seconds. The plot in Figure 8a. shows the balloon prior to the popping event. At this point in time very few events have been recorded. Figure 8b. shows the balloon at the beginning of the popping event. The plot of the fraction of events over time shows a very sharp increase in the slope which indicates the rate of events being recorded is increasing which is commensurate with the fast dynamics associated with a popping balloon. Figure 8c. shows the balloon as the popping event is coming to an end. At this point, the plot of the fraction of events over time exhibits a decrease in slope which is indicative of the slowdown of the dynamics that occurs after the popping event is over. The popping balloon example illustrates how event-based imagers efficiently and adaptive sample data in such a way that the rate of data generation naturally adapts to the rate at which dynamics are evolving as observed in the field of the view of the imager.

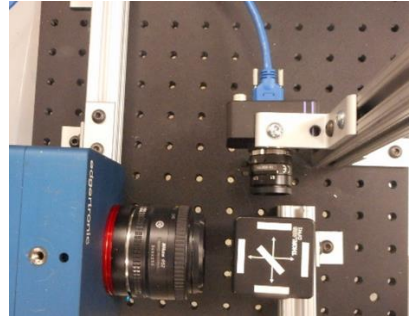

*Figure 7 An event-based imager (top) and conventional high speed imager (left) observing the balloon scene through a 50-50 beam splitter.*

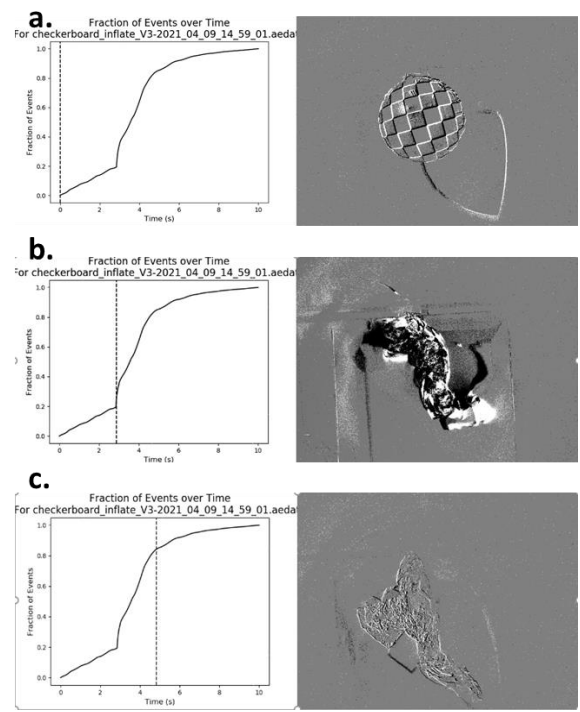

*Figure 8 a.) Plot of the ratio of the events vs time and the event-data associated with the balloon prior to the balloon pop. b.) Plot of the ratio of the events vs time and the event-data associated with the balloon at the time the balloon pops. c.) Plot of the ratio of the events vs time and the event-data associated with the balloon after the balloon pop.*

Next consider the case of observing the laser welding process as shown in Figure 9. The team found that event based imagers resulted in about 35 times less data generation during observation of laser welding melt pools, in comparison to a conventional imager. The reason for the reduction is that an event-based imager is a per-pixel adaptive sampling/detection device and therefore is more data efficient than a conventional imager. In fact, the actual data savings is arguably more efficient than what was originally found because the event based imager does not need additional bytes to record high dynamic range light intensity information. Contemporary event imagers advertize dynamic range of 120 dB [14], which coresponds to 20 bytes. Therefore, the actual memory savings is 2.5 to 3 times more depending on whether the data is store using a conventional 3 byte format. The cost to store event data is not dependent on the dynamic range as is the case for a conventional imager. It is also worth noting that most likely melt-pool imagery of keyholes and their instabilities have a sparse representation over some signal dictionary. Here the term "signal dictionary" refers to a set of signals (i.e. images for this case) that can be linearly combined to form a desired signal/image. A signal dictionary is similar in nature to a basis set, but does not necessarily have the property of being complete and the signals in the dictionary are not necessarily orthogonal. Signal dictionaries can be under-complete or over-complete. As a result it is highly possible that much more signal compression using principle component analysis (PCA) approaches can be achieved from these measurements to further reduce the resources needed for storing and transferring this data. Prior work on unsupervised learning techniques for blind decomposition of structural dynamics using event-based imaging [17], [18] will be applicable to finding even more compressed data representations for in-process monitoring for laser welding.

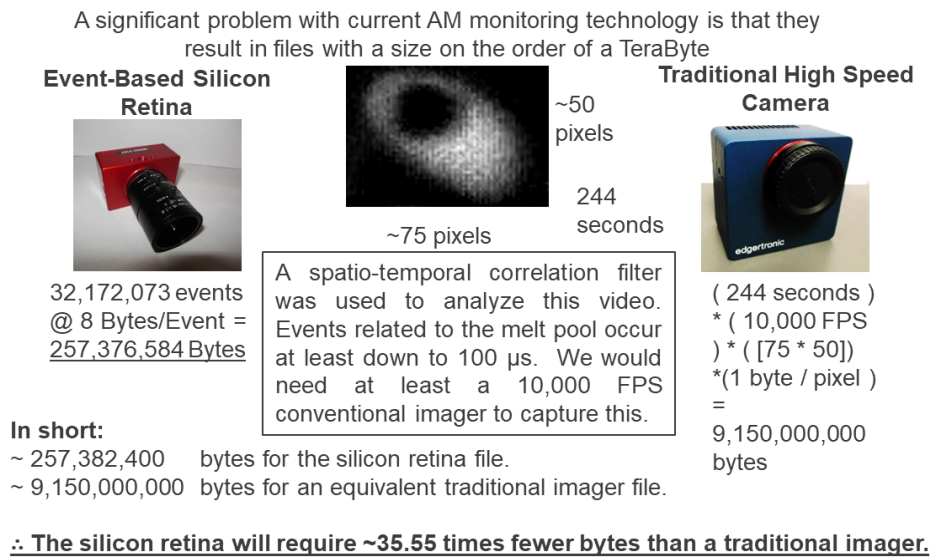

A significant problem with current AM monitoring technology is that they result in files with a size on the order of a TeraByte

**Event-Based Silicon Retina**

~50 pixels

244 seconds

~75 pixels

32,172,073 events @ 8 Bytes/Event = <u>257,376,584 Bytes</u>

A spatio-temporal correlation filter was used to analyze this video. Events related to the melt pool occur at least down to 100 µs. We would need at least a 10,000 FPS conventional imager to capture this.

**Traditional High Speed Camera**

( 244 seconds ) * ( 10,000 FPS ) * ( [75 * 50]) *(1 byte / pixel ) = 9,150,000,000 bytes

**In short:**
~ 257,382,400   bytes for the silicon retina file.
~ 9,150,000,000  bytes for an equivalent traditional imager file.

∴ **The silicon retina will require ~35.55 times fewer bytes than a traditional imager.**

*Figure 9 Analysis of the memory consumption associated with using an event-based imager and a conventional, uniform framerate imager.*

### 3.2 Digital Coded Exposure for Frame Formation

A fundamental question associated with performing event-based processing is how to go about performing data processing on events. Events are a fundamentally different data type when compared to conventional data acquisition approaches based on uniform sampling in time. A result, conventional signal processing tools such as the fast Fourier transform are not immediately applicable. There is currently much interest in moving to fully neuromorphic or event-based paradigms for tasks such as signal processing and learning from event-data. However, this field is still relatively embryonic. For more near term engineering applications hybrid approaches that involve turning event data into conventional frames that can be fed to conventional image processing and machine learning frameworks may be more appropriate. However, the question of how to form frames from event data should not be taken for granted. The most naïve approach to forming frames is to simply sum positive and negative events at each pixel over some defined period in time to generate a frame. This approach to frame formation is shown in *Figure 10*. In this case a video is generated from events with uniform framerate. The videos advance in time by the same time period for each frame, though the time period over which events are aggregated and retained varies between the videos as indicated by the number of microseconds associated with the window. The frames are formed by naïvely summing events over the time period of the frame. The frames from the videos using longer duration windows have greater temporal context; hence the direction of travel for events is more evident in the videos using a longer window duration. It is clear, looking at these images that the length of the window in time, over which events are accumulated, impacts the motion blur properties associated with the resulting image. Recently it was shown in [19] that motion blur has enough degrees-of-freedom that it is possible to generate adversarial examples that resemble motion blur for classification systems. These results indicate that data cleansing/preprocessing of event data when forming frames is important for downstream processing tasks such as classification. To address this issue the concept of the digital coded exposure [20] was created in order to quantitatively control the temporal frequency content of information that contributed to the formation of frames. By quantitatively controlling the frequency content associated with these frames it is more possible to interpret and control the dynamics of an image processing system for online monitoring and control of additive manufacturing.

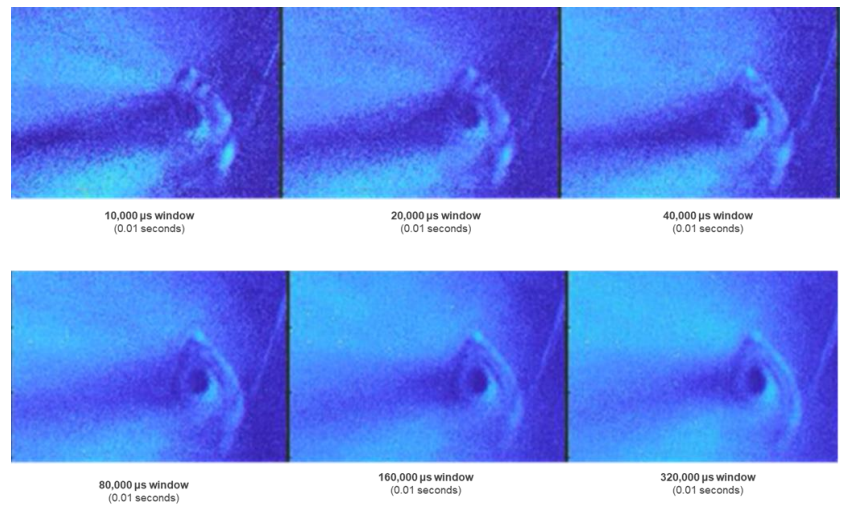

Figure 10 Here we see how the nature of the frame formed from summing events over a sliding window in time changes for different window lengths.

## 5. Conclusions

This work demonstrates that event-based imagers show great potential for in-process monitoring of melt pools associated with welding as well as directed energy deposition additive manufacturing and possibly other forms of additive manufacturing as well. Event-based imagers are seen to be able to observe melt pools created by both lasers and electric arcs without oversaturating their dynamics range when a welding shade is placed in the optical path of the imager. In addition, event-based imagers are shown to have the speed to capture arc welding processes (Figure 4) as well as laser welding processes (Figure 6). Furthermore, we have demonstrated an example where these imagers require 35 times less memory to store data comparable to that of a conventional imager. Event-based imagers have great potential for enabling in-process monitoring and control of welding and additive manufacturing processes as well as for generating digital twins of these processes and the resulting components.

**Author Contributions:** Conceptualization, David Mascareñas; Data curation, David Mascareñas and Andre Green; Formal analysis, David Mascareñas and Andre Green;

Funding acquisition, David Mascareñas, John Bernardin and Garrett Kenyon; Investigation, David Mascareñas, Andre Green, Michael Torrez, Amber Black and Garrett Kenyon; Methodology, David Mascareñas, Andre Green and Ashlee Liao; Project administration, David Mascareñas and Alessandro Cattaneo; Resources, David Mascareñas, Michael Torrez, Amber Black and Garrett Kenyon; Software, David Mascareñas and Andre Green; Validation, David Mascareñas and Andre Green; Visualization, David Mascareñas and Andre Green; Writing – original draft, David Mascareñas, Andre Green and John Bernardin; Writing – review & editing, David Mascareñas, Andre Green, Ashlee Liao, Alessandro Cattaneo, Amber Black and John Bernardin.

**Funding:** This paper was supported by the Laboratory Directed Research and Development program of Los Alamos National Laboratory under project number 20220494MFR, 20220426ER, 20190547MFR as well as other funding sources internal to Los Alamos National Laboratory. Los Alamos National Laboratory is operated by Triad National Security, LLC, for the National Nuclear Security Administration of US Department of Energy (Contract No. 89233218CNA000001).

**Acknowledgments:** The DVS/DAVIS technology was developed by the Sensors group of the Institute of Neuroinformatics (University of Zurich and ETH Zurich), which was funded by the EU FP7 SeeBetter project (grant 270324). This work is a derivative work of a conference proceedings paper [21], and is subject to some copyright restrictions associated with the Society of Experimental Mechanics. Please see:

Mascareñas, D.D., Green, A.W. (2025). Demonstration of Neuromorphic Event-Based Imagers for Optical Measurement of Melt Pools for Additive Manufacturing and Welding Diagnostics. In: Baqersad, J., Di Maio, D., Rohe, D. (eds) Computer Vision & Laser Vibrometry, Vol. 6. IMAC 2024. Conference Proceedings of the Society for Experimental Mechanics Series. Springer, Cham. https://doi.org/10.1007/978-3-031-68192-9_7

A variation of this paper is also published in the Weapons Engineering Symposium and Journal (WESJ) which is not publically accessible.

**Conflicts of Interest:** The authors declare no conflicts of interest. The funders had no role in the design of the study; in the collection, analyses, or interpretation of data; in the writing of the manuscript; or in the decision to publish the results.